\documentclass{emulateapj}
\usepackage{color}
\def\etal{~et al.}
\def\simlt{\lower.5ex\hbox{$\; \buildrel < \over \sim \;$}}
\def\simgt{\lower.5ex\hbox{$\; \buildrel > \over \sim \;$}}

\def\gsim{\lower 2pt \hbox{$\, \buildrel {\scriptstyle >}\over
{\scriptstyle \sim}\,$}}
\def\lsim{\lower 2pt \hbox{$\, \buildrel {\scriptstyle <}\over
{\scriptstyle \sim}\,$}}

\def\deg{\ifmmode ^{\circ}
         \else $^{\circ}$\fi}
\def\pdeg{\ifmmode
           $\setbox0=\hbox{$^{\circ}$}\rlap{\hskip.11\wd0 .}$^{\circ}
     \else \setbox0=\hbox{$^{\circ}$}\rlap{\hskip.11\wd0 .}$^{\circ}$\fi}
     
\def\pc{\ifmmode \mathrm{pc} \else $\mathrm{pc}$ \fi}
\def\mpc{\ifmmode \mathrm{Mpc} \else $\mathrm{Mpc}$\fi}
\def\mpcthree{\ifmmode \mathrm{Mpc}^{-3} \else $\mathrm{Mpc}^{-3}$\fi}
\def\gpcthree{\ifmmode \mathrm{Gpc}^{-3} \else $\mathrm{Gpc}^{-3}$\fi}

\def\kelvin{\ifmmode \mathrm{K} \else {$\mathrm{K}$}\fi}
\def\kev{\ifmmode \mathrm{keV} \else $\mathrm{keV}$ \fi}

\def\lsun{\ifmmode {L_\odot} \else $L_\odot$\fi}
\def\msun{\ifmmode M_\odot \else $M_\odot$\fi}
\def\msunyr{\ifmmode M_\odot~\mathrm{yr}^{-1} \else $M_\odot~\mathrm{yr}^{-1}$\fi}

\def\cosi{\ifmmode {\cos\,i} \else $\cos\,i$\fi}

\def\heii{\ifmmode {\rm He{\sc ii}} \else He~{\sc ii}\fi}
\def\mgii{\ifmmode {\rm Mg{\sc ii}} \else Mg~{\sc ii}\fi}
\def\caii{\ifmmode {\rm Ca{\sc ii}} \else Ca~{\sc ii}\fi}
\def\ciii{\ifmmode {\rm C{\sc iii}]} \else C~{\sc iii}]\fi}
\def\civ{\ifmmode {\rm C{\sc iv}} \else C~{\sc iv}\fi}
\def\mgii{\ifmmode {\rm Mg{\sc ii}} \else Mg~{\sc ii}\fi}

\newcommand{\oiii}{{\sc [O~iii]}}

\newcommand{\nev}{{[Ne~{\sc v}]}}

\newcommand{\fevii}{[Fe~{\sc vii}]}

\def\teff{\ifmmode {T_{\rm eff}} \else $T_{\rm eff}$\fi}
\def\tmax{\ifmmode {T_{\rm max}} \else $T_{\rm max}$\fi}

\def\mbh{\ifmmode {M_{\rm BH}} \else $M_{\rm BH}$\fi}
\def\led{\ifmmode L_{\mathrm{Ed}} \else $L_{\mathrm{Ed}}$\fi}
\def\lbolflare{\ifmmode L_{\mathrm{bol,flare}} \else $L_{\mathrm{bol,flare}}$\fi}
\def\lagn{\ifmmode L_{\mathrm{agn}} \else $L_{\mathrm{agn}}$\fi}
\def\lbolagn{\ifmmode L_{\mathrm{bol,agn}} \else $L_{\mathrm{bol,agn}}$\fi}
\def\lbol{\ifmmode L_{\mathrm{bol}} \else $L_{\mathrm{bol}}$\fi}
\def\mdot{\ifmmode {\dot M} \else $\dot M$\fi}
\def\mdoto{\ifmmode {\dot{M}_0} \else  $\dot{M}_0$\fi}
\def\mdotf{\ifmmode {\dot{M}_\mathrm{flare}} \else  $\dot{M}_\mathrm{flare}$\fi}

\def\hnot{\ifmmode H_0 \else H$_0$ \fi}

\def\vkep{\ifmmode v_\mathrm{Kep} \else $v_\mathrm{Kep}$ \fi}
\def\vc{\ifmmode v_\mathrm{c} \else $v_\mathrm{c}$ \fi}

\def\vthree{\ifmmode v_{1000} \else $v_{1000}$ \fi}
\def\vrel{\ifmmode v_\mathrm{rel} \else $v_\mathrm{rel}$ \fi}
\def\vkick{\ifmmode v_\mathrm{kick} \else $v_\mathrm{kick}$ \fi}
\def\vkickz{\ifmmode v_{\mathrm{kick},z} \else $v_{\mathrm{kick},z} $ \fi}
\def\vkicky{\ifmmode v_{\mathrm{kick},y} \else $v_{\mathrm{kick},y} $ \fi}
\def\vchar{\ifmmode v_\mathrm{char} \else $v_\mathrm{char}$ \fi}
\def\eflare{\ifmmode E_\mathrm{flare} \else $E_\mathrm{flare}$ \fi}
\def\ekick{\ifmmode E_\mathrm{kick} \else $E_\mathrm{kick}$ \fi}
\def\ecoll{\ifmmode E_\mathrm{coll} \else $E_\mathrm{coll}$ \fi}
\def\ezero{\ifmmode E_\mathrm{0} \else $E_\mathrm{0}$ \fi}
\def\efac{\ifmmode \xi_\mathrm{E} \else $\xi_\mathrm{E}$ \fi}
\def\tqso{\ifmmode t_\mathrm{QSO} \else $t_\mathrm{QSO}$ \fi}
\def\tflare{\ifmmode t_\mathrm{flare} \else $t_\mathrm{flare}$ \fi}
\def\tzero{\ifmmode t_\mathrm{0} \else $t_\mathrm{0}$ \fi}
\def\tfac{\ifmmode \xi_\mathrm{t} \else $\xi_\mathrm{t}$ \fi}
\def\gfac{\ifmmode f_\mathrm{g} \else $f_\mathrm{g}$ \fi}
\def\lflare{\ifmmode L_\mathrm{flare} \else $L_\mathrm{flare}$ \fi}
\def\fflare{\ifmmode F_\mathrm{flare} \else $F_\mathrm{flare}$ \fi}
\def\nflare{\ifmmode N_\mathrm{flare} \else $N_\mathrm{flare}$ \fi}
\def\tshock{\ifmmode T_\mathrm{shock} \else $T_\mathrm{shock}$ \fi}
\def\rmin{\ifmmode R_\mathrm{1} \else $R_\mathrm{1}$ \fi}
\def\rmax{\ifmmode R_\mathrm{2} \else $R_\mathrm{2}$ \fi}
\def\rbound{\ifmmode R_\mathrm{b} \else $R_\mathrm{b}$ \fi}
\def\pbound{\ifmmode P_\mathrm{b} \else $P_\mathrm{b}$ \fi}
\def\mbound{\ifmmode M_\mathrm{b} \else $M_\mathrm{b}$ \fi}
\def\mbo{\ifmmode M_{\mathrm{b}0} \else $M_{\mathrm{b}0} $ \fi}
\def\ebo{\ifmmode E_{\mathrm{b}0} \else $E_{\mathrm{b}0} $ \fi}
\def\efinal{\ifmmode E_\mathrm{final} \else $E_\mathrm{final} $ \fi}
\def\tbound{\ifmmode t_\mathrm{b} \else $t_\mathrm{b}$ \fi}
\def\tagn{\ifmmode t_\mathrm{AGN} \else $t_\mathrm{AGN}$ \fi}
\def\torb{\ifmmode t_\mathrm{orb} \else $t_\mathrm{orb}$ \fi}
\def\tdf{\ifmmode t_\mathrm{df} \else $t_\mathrm{df}$ \fi}
\def\rlim{\ifmmode R_\mathrm{lim} \else $R_\mathrm{lim}$ \fi}
\def\vlim{\ifmmode v_\mathrm{lim} \else $v_\mathrm{lim}$ \fi}
\def\vphi{\ifmmode v_\phi \else $v_\phi$ \fi}
\def\mlim{\ifmmode M_\mathrm{lim} \else $M_\mathrm{lim}$ \fi}
\def\tlim{\ifmmode t_\mathrm{lim} \else $t_\mathrm{lim}$ \fi}
\def\llim{\ifmmode L_\mathrm{lim} \else $L_\mathrm{lim}$ \fi}
\def\fqso{\ifmmode f_\mathrm{QSO} \else $f_\mathrm{QSO}$ \fi}

\def\hbeta{\ifmmode \rm{H}\beta \else H$\beta$\fi}
\def\hbetan{\ifmmode \rm{H}\beta_{\rm n} \else H$\beta_{\rm n}$\fi}
\def\hgamma{\ifmmode \rm{H}\gamma \else H$\gamma$\fi}
\def\hdelta{\ifmmode \rm{H}\delta \else H$\delta$\fi}
\def\hepsilon{\ifmmode \rm{H}\epsilon \else H$\epsilon$\fi}
\def\hzeta{\ifmmode \rm{H}\zeta \else H$\zeta$\fi}
\def\halpha{\ifmmode \rm{H}\alpha \else H$\alpha$\fi}
\def\lalpha{\ifmmode \rm{Ly}\alpha \else Ly$\alpha$}

\def\dvhb{\ifmmode \Delta v_{\hbeta} \else $\Delta v_{\hbeta}$\fi}
\def\dvmg{\ifmmode \Delta v_{\rm{Mg}} \else $\Delta v_{\rm{Mg}}$\fi}

\def\muobs{\ifmmode {\mu_{o}} \else  $\mu_{o}$ \fi}
\def\cosi{\ifmmode {\mathrm{cos}\,i} \else $\mathrm{cos}\,i$\fi}

\def\teff{\ifmmode {T_{eff}} \else $T_{eff}$ \fi}
\def\tmax{\ifmmode {T_{max}} \else $T_{max}$ \fi}

\def\tauh{\ifmmode {\tau_{\rm H}} \else $\tau_{\rm H}$ \fi}

\def\yr{\ifmmode {\rm yr} \else  yr \fi}
\def\kms{\ifmmode \rm km~s^{-1}\else $\rm km~s^{-1}$\fi}
\def\cm{\ifmmode {\rm cm} \else  cm \fi}
\def\cmmitwo{\ifmmode \rm cm^{-2} \else $\rm cm^{-2}$\fi}
\def\cmmithree{\ifmmode \rm cm^{-3} \else $\rm cm^{-3}$\fi}
\def\cmps{\ifmmode \rm cm~s^{-1}\else $\rm cm~s^{-1}$\fi}
\def\cmpsps{\ifmmode \rm cm~s^{-2}\else $\rm cm~s^{-2}$\fi}
\def\kmps{\ifmmode \rm km~s^{-1}\else $\rm km~s^{-1}$\fi}
\def\kmpspmpc{\ifmmode \rm km~s^{-1}~Mpc^{-1} \else
    $\rm km~s^{-1}~Mpc^{-1}$\fi}
  
\def\gcmthree{\ifmmode \rm g~cm^{-3} \else $\rm g~cm^{-3}$\fi}
\def\gcmtwo{\ifmmode \rm g~cm^{-2} \else $\rm g~cm^{-2}$\fi}
   
\def\erg{\ifmmode {\rm erg} \else $\rm erg$ \fi}
\def\ergps{\ifmmode {\rm erg~s^{-1}} \else $\rm erg~s^{-1}$ \fi}
\def\ergcms{\ifmmode \rm erg~cm^{-2}~s^{-1} \else $\rm erg~cm^{-2}~s^{-1}$ \fi}
\def\ergcmshz{\ifmmode \rm erg~s^{-1}~cm^{-2}~Hz^{-1} \else $\rm
erg~cm^{-2}~s^{-1}~Hz^{-1}$ \fi}
\def\ergcmsa{\ifmmode \rm erg~cm^{-2}~s^{-1}~\AA^{-1} \else $\rm
erg~cm^{-2}~s^{-1}~\AA^{-1}$ \fi}
\def\ergshz{\ifmmode \rm erg s^{-1} Hz^{-1} \else
   $\rm erg s^{-1} Hz^{-1}$ \fi}

\def\lam{\ifmmode {\lambda} \else {$\lambda$} \fi}
\def\llam{\ifmmode {L_\lambda} \else  $L_\lambda$ \fi}
\def\lamLlam{\ifmmode \lambda L_{\lambda}(5100) \else {$\lambda L_{\lambda}(5100)$} \fi}
\def\nuLnu{\ifmmode \nu L_{\nu}(5100) \else {$\nu L_{\nu}(5100)$} \fi}
\def\ilam{\ifmmode {I_\lambda} \else  $I_\lambda$ \fi}
\def\flam{\ifmmode {F_\lambda} \else  $F_\lambda$ \fi}
\def\inu{\ifmmode {I_\nu} \else  $I_\nu$ \fi}
\def\fnu{\ifmmode {F_\nu} \else  $F_\nu$ \fi}
\def\bnu{\ifmmode {B_\nu} \else  $B_\nu$ \fi}

\def\msigma{\ifmmode M_{\sigma} \else $M_{\sigma}$\fi}
\def\mbulge{\ifmmode M_{\mathrm{bulge}} \else $M_{\mathrm{bulge}}$\fi}
\def\mgal{\ifmmode M_{\mathrm{gal}} \else $M_{\mathrm{gal}}$\fi}
\def\lgal{\ifmmode L_{\mathrm{gal}} \else $L_{\mathrm{gal}}$\fi}
\def\lbulge{\ifmmode L_{\mathrm{bulge}} \else $L_{\mathrm{bulge}}$\fi}
\def\mgalstar{\ifmmode M^*_{\mathrm{gal}} \else $M^*_{\mathrm{gal}}$\fi}

\def\mbhsigstar{\ifmmode M_{\mathrm{BH}} - \sigma_* \else $M_{\mathrm{BH}} - \sigma_*$ \fi}
\def\deltalogmbh{\ifmmode \Delta~{\mathrm{log}}~M_{\mathrm{BH}} \else $\Delta$~log~$M_{\mathrm{BH}}$\fi}

\def\sigstar{\ifmmode \sigma_* \else $\sigma_*$\fi}
\def\sigthree{\ifmmode \sigma_{\mathrm{[O~III]}} \else $\sigma_{\mathrm{[O~III]}}$\fi}
\def\sigtwo{\ifmmode \sigma_{\mathrm{[O~II]}} \else $\sigma_{\mathrm{[O~II]}}$\fi}
\def\signl{\ifmmode \sigma_{\mathrm{NL}} \else $\sigma_{\mathrm{NL}}$\fi}
\def\wthree{\ifmmode {\rm FWHM({[O~III]})} \else $FWHM({[O~III]})$ \fi}
\def\wtwo{\ifmmode {\rm FWHM({[O~II]})} \else $FWHM({[O~II]})$ \fi}
\def\mthree{\ifmmode M_{\mathrm [O~III]} \else $M_{\mathrm [O~III]}$ \fi}
\def\mtwo{\ifmmode M_{\mathrm [O II]} \else $M_{\mathrm [O II]}$ \fi}

\def\lbreak{\ifmmode L_{\mathrm{break}} \else $L_{\mathrm{break}}$\fi}
\def\lcut{\ifmmode L_{\mathrm{cut}} \else $L_{\mathrm{cut}}$\fi}

\slugcomment{Submitted to ApJ}

\shortauthors{Smith, Shields, Salviander, Stevens \& Rosario}
\shorttitle{Equal Peaked AGN}

\begin{document}

\title{Double-Peaked Narrow-Line AGN II. The Case of Equal Peaks}

\author{K.~L. Smith\altaffilmark{1}, G.~A. Shields\altaffilmark{1}, S.~Salviander\altaffilmark{1,2}, A.~C. Stevens\altaffilmark{1} \& D.J. Rosario\altaffilmark{3}}

\altaffiltext{1}{Department of Astronomy, University of Texas, Austin,
TX 78712; klsmith@astro.umd.edu, shields@astro.as.utexas.edu, triples@astro.as.utexas.edu, acs0196@mail.utexas.edu} 

\altaffiltext{2}{Department of Physics, Southwestern University, Georgetown, TX 78626}

\altaffiltext{3}{Max-Planck-Institute for Extraterrestrial Physics, Garching, 85748; rosario@mpe.mpg.de}

\begin{abstract}

AGN with double-peaked narrow lines (DPAGN) may be caused by kiloparsec scale binary AGN, bipolar outflows, or rotating gaseous disks. We examine the class of DPAGN in which the two narrow line components have closely similar intensity as being especially likely to involve disks or jets.   Two spectroscopic indicators support this likelihood.   For DPAGN from \citet{smith10}, the ``equal-peaked'' objects (EPAGN)  have \nev/\oiii ratios lower than for a control sample of non-double peaked AGN.  This  is unexpected for a pair of normal AGN in a galactic merger, but may be consistent with \oiii\ emission from a rotating ring with relatively little gas at small radii.  Also,  \oiii/\hbeta\ ratios of the redshifted and blueshifted systems in the EPAGN are more similar to each other than in a control sample, suggestive of a single ionizing source and inconsistent with the binary interpretation.

\end{abstract}


\section{Introduction}
\label{sec:intro}

 Dual AGN have recently attracted interest as examples of black hole fueling and AGN visibility during galaxy mergers.  Several recent studies have considered objects with double-peaked narrow emission lines, particularly the \oiii~$\lambda\lambda~5007,4959$ line, as candidates for kiloparsec-scale binary AGN \citep{zhou04, gerke07, comerford09a, wang09, liu10a, smith10}. Two adjacent narrow-line regions (NLRs) associated with individual SMBHs would result in a double-peaked profile for plausible line-of-sight orbital velocities. Recent imaging of double-peaked narrow line AGN (hereinafter ``DPAGN'') has resulted in the discovery of a number of true binaries.   \citet{liu10b} report four such objects, with dual active nuclei in optical slit spectra spatially coinciding with NIR-detected double stellar components. Also, \citet{fu11a} report that 8 out of 17 observed Type 1 DPAGN and 8 out of 33 observed Type 2 DPAGN were found to be in mergers, although not necessarily in mergers where the orbital velocity between the galaxies is responsible for the observed line splitting. \citet{shen10} present five DPAGN imaged to be kiloparsec-scale binaries, and \citet{mcgurk11} have reported a confirmed  dual AGN in J0952+2552.  \citet{comerford11}  report a dual AGN selected by double-peaked \oiii\ as a \emph{Chandra} double X-ray source.  

Double peaked narrow lines can also result from biconical outflow, as in Mrk 78 \citep{whittle04} and in J1517+33 and J1129+60 \citep{rosario10}, or from a rotating gaseous disk.   \citet{shen10} maintain that the majority (up to 90\%) of double-peaked \oiii\ objects are caused by NLR kinematics (i.e., jets and disks) instead of kpc-scale binaries.   \citet{fu11b}  find only 4 binary AGN among 42 DPAGN.
A recent study by \citet{tingay11} sought binary AGN by looking for dual compact radio sources in 11 of the double-peaked \oiii\ objects published by \citet{wang09}. These authors find compact radio emission in only two of their sources, with neither detection being a dual source.  In a recent theoretical study, \citet{blecha12} examine narrow line profiles in galaxy mergers with fueling of the two central black holes.  They find that double-peaked profiles often result from gas dynamics distinct from binarity of the active nuclei, but that double-peaked \oiii\ lines still represent a useful search method for finding dual AGN.  It is important to develop diagnostic tools to identify the likely cause of double peaked narrow lines in a given AGN, in order to improve efficiency of finding dual AGN as well as to promote understanding of the range of geometries that occur in the narrow line region (NLR) of AGN.

In this paper, we focus on DPAGN in which the two components of the narrow lines have substantially equal strength (``equal-peaked AGN" or ``EPAGN") .  This is motivated in particular by the existence of a subset of DPAGN with strikingly symmetrical profile, which bring to mind the possibility of a rotating ring or disk as the cause of the double peak.  Such objects might also represent biconical outflows, but they seem unlikely to be dual AGN, for which there is no reason to expect equal component fluxes.  In Section \ref{sec:selection}, we describe how these objects were selected from the larger sample of DPAGN. In Section \ref{sec:linerat}, we look at the high-ionization line ratios of the EPAGN, and compare the line ratios in the red- and blue-shifted systems. We find that these properties of the EPAGN as a group are inconsistent with a binary interpretation. The properties we find are consistent with a rotating-disk interpretation for the double-peaked \oiii\ profile, but stress that this interpretation is not unique.
Our conclusions are summarized in Section \ref{sec:conclusion}.
We assume a cosmology with 
$\Omega_{\Lambda}=0.7$, $\Omega_{m}=0.3$, and H$_{0}=70$ \kms Mpc$^{-1}$.

\section{Selection of Equal-Peaked Objects}
\label{sec:selection}

The original selection of the double-peaked \oiii\ objects is described in Section 4 of \citet{smith10}. 
The objects were selected from SDSS DR7, which contains 21,592 QSO spectra in the redshift range of $0.1\leq z\leq 0.7$.  This keeps the \oiii~$\lambda\lambda~5007,4959$ line in the SDSS spectral window and out of the water vapor forest.  A visual inspection found 148 DPAGN, of which 62 were Type 2 and 86 Type 1, with an average redshift of 0.33.

\begin{figure}[ht]
\begin{center}
\plotone{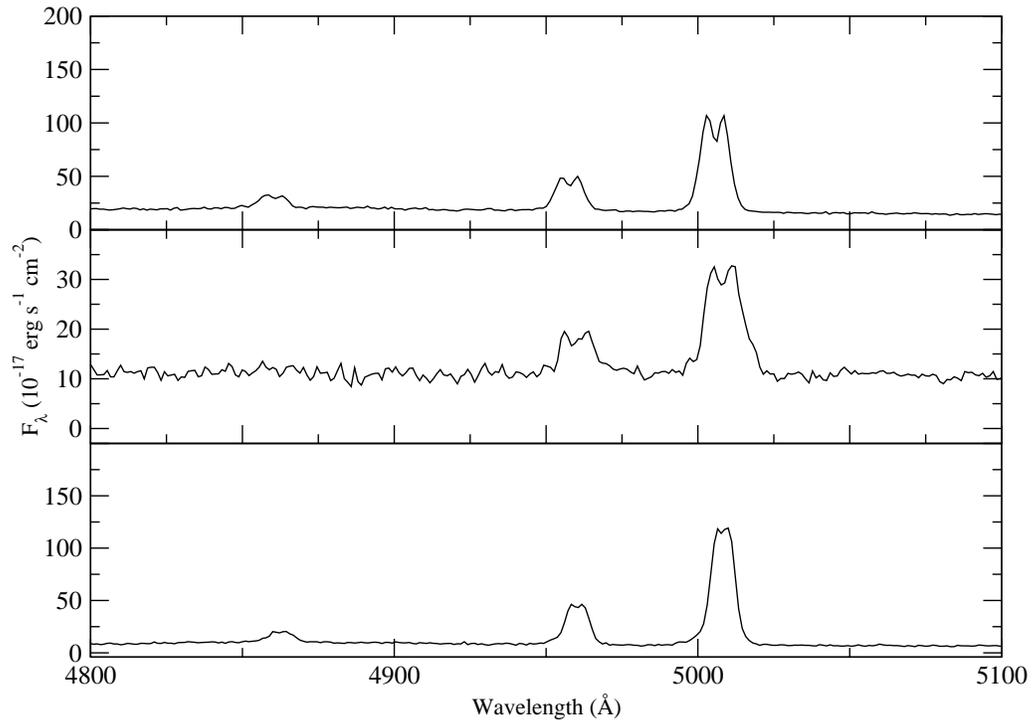}
\figcaption[]{Three spectra typical of the eye-selected equal-peaked objects (EHAGN). Note the similarity in height and structure of the two components of \oiii~$\lambda\lambda~5007,4959$.
\label{fig:examples} }
\end{center}
\end{figure}

We used two different methods to identify the equal-peaked objects. The first method is selection by eye. Symmetrical appearance was the basis for this selection; i.e., the red and blue components are of similar width and height. Several of these objects are ``dimples," where the components are separated by only a small depression at the top of the line profile. (It is these that we most suspect of being rotating disks.)  Others are more widely spaced and have a deeper gap between the peaks, but otherwise are similar in shape. Examples are shown in Figure \ref{fig:examples}. We identified thirteen such objects out of the 148 total DPAGN. Henceforth, objects selected in this fashion are referred to as ``equal-height AGN" or EHAGN.  The objects thus selected are listed Table \ref{t:tab1}; there are 6 EHAGN AGN of Type 1 and 7 of Type 2.

A more quantitative way to identify equal-peaked objects is by using the ratio of the component fluxes. In Table 3 of \citet{smith10}, we reported the component fluxes for 78 objects for which we obtained reliable double Gaussian fits. The fitting procedure assumed a Gaussian for each of the two components of the $\lambda5007$ and $\lambda4959$ lines, with adjustable velocity, FWHM, and flux, while maintaining the 3:1 flux ratio required by atomic physics. From these fluxes, we construct the ratio $F_r/F_b$, where $F_r$ and $F_b$ are the fluxes of the red and blue components, respectively. We take objects with $0.75\leq F_r/F_b\leq 1.25$ to be ``equal'' and refer to them as ``equal-flux AGN" or EFAGN.  There are 13 EFAGN AGN of Type 1 and 9 of Type 2, as listed in Table \ref{t:tab2}.  Note that these lists are not mutually exclusive.  There are 3 Type 1 and 2 Type 2 objects in common.

Some of the EHAGN did not have Gaussian fits we considered to be unique.  Also, several of the EFAGN were unequal to the eye and seem unlikely to be caused by a rotating disk unless it is eccentric or subject to asymmetrical obscuration.  For these reasons, we analyze the EHAGN and EPAGN separately, and emphasize the former.
 
\citet{shen10} describe imaging and spatially resolved spectroscopy of a sample of DPAGN from SDSS.  Many of the objects are argued to be single AGN with complex NLR kinematics, based on an absence of dual stellar nuclei and the nature of the \oiii\ velocity field.  These authors identify ten double-peaked \oiii\ objects as likely to be caused by rotating gaseous disks. Five of these objects would have been classified as EHAGN in our approach. The remaining five objects appear to be EFAGN, with one exception, J0851+1327, whose \oiii\ line exhibits only a slight blue shoulder. Additionally, one of our EHAGN and three of our EFAGN appear in the imaged sample of \citet{fu11b}, where the authors conclude from high resolution imaging and spectroscopy that the double-peaked \oiii\ results from NLR kinematics. Therefore, examples already exist in which EPAGN have been determined to be caused by NLR kinematics, likely rotating disks, instead of binarity.

\section{The Nature of the EPAGN}
\label{sec:linerat}

Here we consider two spectroscopic tests of the nature of the EPAGN.

\subsection{High-Ionization Lines}
\label{sec:highion}

One possible test of the nature of the EPAGN involves the high ionization narrow lines.  It is widely believed that much of the emission in high ionization lines such as \nev\ and \fevii\  comes from the core of the NLR, because they often have larger line widths and because they have weaker intensities in Type 2 AGN (where the NLR core may be obscured.)  In the case of a rotating ring, the emission from such a core might be relatively weak.  Thus, weakness of \nev\ in the EPAGN would be a  point of consistency with a rotating disk geometry, although this would not be a unique interpretation.  Such weakness would in any case argue against a dual AGN model, which presumably would involve two normal NLRs.

We examined the \nev\ intensity of the EPAGN, restricting attention to the Type 1 objects for which the NLR core should be visible.  For the EHAGN, there are 6 objects (see Table \ref{t:tab1}), of which 4 have redshifts large enough for \nev\ to be in the spectral window.   For these, line fluxes were measured using the IRAF routine SPLOT\footnote{IRAF is distributed by the National Optical Astronomy Observatories, which are operated by the Association of Universities for Research in Astronomy, Inc., under cooperative agreement with the National Science Foundation.}.  For J1131+61, \oiii\ is weak and \nev\ undetected.  For the remaining three objects, the measured fluxes for \nev\ and \oiii\ are given in Table \ref{t:tab3}.  
 We estimate an uncertainty of $\pm10\%$ in the individual measurements of the flux in \nev, based on the highest and lowest plausible continuum levels. The results give $I(\lambda3426)/I(\lambda5007)$ in the range 0.056 to 0.075, for an average of $0.064\pm0.005$ (standard error of mean). 
 
 For comparison, the \nev/\oiii\ ratio in the overall SDSS Type 1 AGN composite by \citet{vdb01} is 0.21.  We made our own composite spectrum for a control sample of all SDSS AGN in the redshift range $0.2 < z < 0.5$, corresponding to our EHAGN having \nev\ in the spectral window . [This sample has 3346 objects, overwhelmingly Type 1. To create the composite, the individual spectra were normalized to an average flux \flam\ of unity across the entire spectrum, and then averaged together, as described by \citet{salviander07}.] This composite has $I(\lambda3426)/I(\lambda5007) = 0.19\pm0.02$, where the error reflects uncertainty in the continuum level at \nev.  Thus, the group of three EHAGN has substantially weaker \nev\ than is typical for Type 1 AGN in SDSS.   We also examined the distribution of \nev\ intensities in the control sample in order to assess the likelihood of getting small values for the three EHAGN by chance.   We selected the first 100 SDSS Type 1 AGN  at $z > 0.5$ for measurement of \oiii\ and \nev\ with SPLOT.  Of the 27 objects for which \nev\ could reliably be measured, only 4 have \nev/\oiii\ less than 0.08, giving a probability of less than 1\%\ that all three EHAGN would be as low as observed by chance.  

For the EFAGN, there are 11 objects of Type 1 excluding those that are already in the EHAGN sample.   Of these, 6 objects allowed a reliable measurement of \nev.  The resulting values of \nev/\oiii\ ranged from 0.049 to 0.092, with an average of  $0.069\pm0.006$ (error of mean).   These values are again lower than the typical value for the single-peak control sample.

We conclude that the EPAGN do have exceptionally weak \nev.   This implies  the EPAGN for the most part are not binaries.  The weak \nev\  implies that the NLR of the EPAGN is different from normal AGN in some way that weakens the high ionization lines.  This is consistent with a rotating ring, but other models involving NLR geometry are possible.    We caution that the sample is small, but the potential utility of this approach is evident.
 
\subsection{\oiii\ / \hbeta\ in Equal-Peaked Objects}
\label{sec:hbeta}

For dual AGN, the red and blue components of the narrow lines are assumed to be due to the presence of two separate NLRs, and the line ratios in the blueshifted system would in general not be equal to those of the redshifted system.  For a rotating disk model, however, a single ionizing source is responsible for illuminating both systems. Therefore, equality of line ratios in the two velocity components may provide a test of the nature of the double peaks.  This idea is mentioned by \citet{xu09} when considering a possible rotating disk explanation for the double-peaked object SDSS J1316+1753, and by \citet{liu10a} in the original presentation of their double-peaked sample.

We have performed double Gaussian fits for the \hbeta\ lines in our EPAGN and for the other DPAGN from \citet{smith10} that exhibited double-peaked \hbeta. Only Type 2 objects are considered, as separation of the broad and narrow components of \hbeta\ in Type 1 objects proved impractical. There are 5 EHAGN and 5 EFAGN usable for this purpose.  From these results and the double Gaussian fits of the \oiii\ $\lambda5007$ lines mentioned above, we calculate the \oiii/\hbeta\ ratio for the red- and blueshifted systems of each object, and compare them.  We create the ratio of ratios $R_r/R_b$, where $R_r=$\oiii$_r$/\hbeta$_r$ is the line ratio for the redshifted system and likewise for the blueshifted system. As a measure of how similar these ratios are for a given object, we use the value $|1-R_r/R_b|$.  For the EHAGN and the EFAGN samples, we obtain average values \hbox{$\langle|1-R_r/R_b|\rangle =0.07\pm0.03$} and $0.13\pm0.09$ (error of mean), respectively.

How do these values compare with what would be expected for dual AGN?  We formed a control sample of 25 Type 2 AGN selected by eye from the parent sample of SDSS AGN.  We measured the \oiii\ and \hbeta\ line fluxes and calculated the \oiii\ / \hbeta\ line ratios for these objects.   These ratios were then randomly combined with one another in pairs to examine the distribution of values of $\langle|1-R_r/R_b|\rangle$.  We obtained an average value of \hbox{$\langle|1-R_r/R_b|\rangle =0.52\pm0.13$} (error of the mean).   This exceeds the value for the EHAGN by more than $3\sigma$ and for the EFAGN by more
than $2\sigma$.  Thus, the line ratios of the red- and blueshifted systems are more similar to one another among the EPAGN than expected for random pairings of normal quasars.  This is suggestive of a single ionizing source and inconsistent with binaries.

As a further test of the statistical significance of this result, we created an independent set of 10 random pairings of the list of 25 Type 2 single-peak sample with itself, giving 10 sets of 25 simulated binaries. For each of these 10 pairings, we selected at random 72 subsets of 5 objects, giving 720 such subsets. For these subsets of 5 simulated objects, the average value of \hbox{$\langle|1-R_r/R_b|\rangle$} was 0.51, with an rms scatter of 0.10.  The minimum value was 0.10, and 12 cases had \hbox{$\langle|1-R_r/R_b|\rangle < 0.20$}. We conclude that a result of 0.07 (for the EHAGN) or 0.12 (for the EFAGN) will occur by chance with a probability of $\lsim1\%$.  

\begin{figure}[ht]
\begin{center}
\plotone{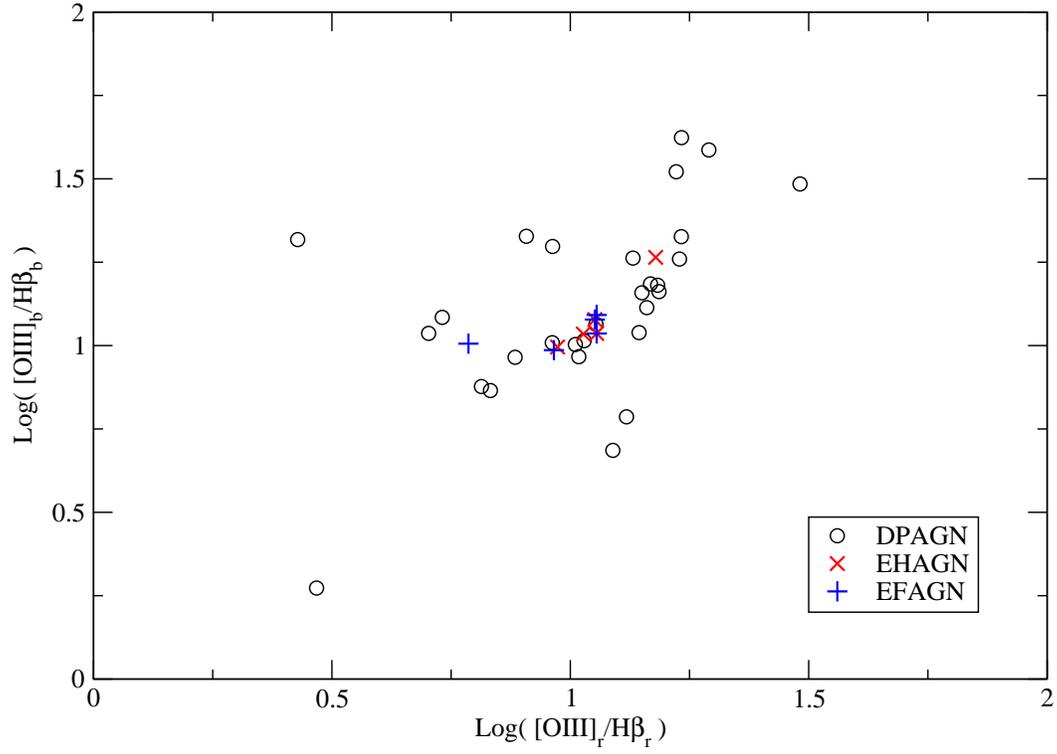}
\figcaption[]{Ratios of the \oiii/\hbeta\ lines for the blueshifted systems of the Type 2 DPAGN to those of the redshifted systems. A correlation exists similar to the one reported in \citet{liu10a}. Hollow circles represent the overall DPAGN, red crosses represent the eye-selected equal height AGN (EHAGN), and blue plus-signs represent the equal flux AGN (EFAGN). The EPAGN appear to occupy a tighter 1:1 correlation than the overall sample, consistent with rotating disks.
\label{fig:hbeta}}
\end{center}
\end{figure}

For the Type 2  DPAGN sample as a whole, excluding the equal-peaked objects, we find $\langle|1-R_r/R_b|\rangle = 0.26\pm0.09$ (error of the mean).  This value is intermediate between the EPAGN and the control sample, but the difference is only marginally significant in therms of the error bars.   In Figure \ref{fig:hbeta}, we plot the values of the red- and blue-shifted systems against one another. It can be seen that these values are correlated in the same manner as in Figure 4 of \citet{liu10a}. The Spearman correlation coefficient for our entire Type 2 DPAGN sample is $\rho=0.60$, with a probability of null correlation $P_0=9.52\times10^{-4}$. The number of EPAGN is too small to form such a statistic.   However, it appears that the EPAGN are less scattered and lie nearer to the 1:1 line than the DPAGN in general.  The intermediate degree of correlation of $R_r$ with $R_b$ for the general DPAGN sample is consistent with a mix of binaries, disks, and jets; but such an interpretation may not be unique.

\section{Conclusion}
\label{sec:conclusion}

AGN with double-peaked narrow emission lines are attracting increased attention as candidates for dual AGN.  
DPAGN offer potential insights into several aspects of AGN physics.   If caused by binary AGN, they offer a way to assess the probability of simultaneous fueling of both black holes in a galactic merger.  If caused by rotating disks or radio jets, they offer insights into NLR geometry and gas dynamics.   Radio, X-ray, and optical imaging, together with spatially resolved spectroscopy, are required to take advantage of these opportunities. 

In this paper, we have focused on the case in which the two peaks of the \oiii\ line have nearly equal intensities.  Such objects have weaker high-ionization lines than a control sample of single-peaked Type 1 AGN, making them unlikely to be binary quasars.  Line ratios for the red- and blueshifted systems of the EPAGN are more similar to each other than those of the overall DPAGN sample, which in turn are more similar to each other than for a randomly combined control sample of single-peaked AGN.  These findings suggest a single ionizing source and are inconsistent with a binary scenario.  It is plausible that such profiles often represent rotating disks, but further study is required to confirm this conclusion.

\acknowledgments

We thank Julie Comerford, Jenny Greene, Mark Whittle, and Bev Wills for helpful discussions.   We thank an anonymous referee for many helpful suggestions that led to significant improvements in the manuscript.  KLS and GAS  gratefully acknowledge support from the University Cooperative Society of the University of Texas at Austin and the Jane and Roland Blumberg Cenntenial Professorship in Astronomy.

Funding for the Sloan Digital Sky Survey (SDSS) has been provided by the Alfred P. Sloan Foundation, the Participating Institutions, the National Aeronautics and Space Administration, the National Science Foundation, the U.S. Department of Energy, the Japanese Monbukagakusho, and the Max Planck Society. The SDSS Web site is http://www.sdss.org/. The SDSS is managed by the Astrophysical Research Consortium (ARC) for the Participating Institutions. The Participating Institutions are The University of Chicago, Fermilab, the Institute for Advanced Study, the Japan Participation Group, The Johns Hopkins University, the Korean Scientist Group, Los Alamos National Laboratory, the Max-Planck-Institute for Astronomy (MPIA), the Max-Planck-Institute for Astrophysics (MPA), New Mexico State University, University of Pittsburgh, University of Portsmouth, Princeton University, the United States Naval Observatory, and the University of Washington.

\clearpage

%=========Table 1================

\begin{deluxetable}{lcccc}
\label{tab:list}
\tablewidth{0pt}
\tablecaption{Equal Height  Double-Peaked AGN\label{t:tab1}}
\tablehead{
\colhead{SDSS Name}          &
\colhead{Velocity Splitting}       &
\colhead{$z_{\rm SDSS}$} &
\colhead{$F_{\rm r}/F_{\rm b}$}            &
\colhead{Seyfert Type} \\
\colhead{ }  &
\colhead{km/s} &
\colhead{ } &
\colhead{ } &
\colhead{ }}
\startdata

J101241.20+215556.0	&	210	&	0.111	&	1.027	&	1	\\
J113105.07+610405.1 	&	330	&	0.338	&			&	1	\\
J131018.47+250329.5	&	180	&	0.313	&	0.723	&	1	\\
J144105.64+180507.9	&	280	&	0.107	&			&	1	\\
J151518.29+551535.3 	&	250	&	0.513	&	1.214	&	1	\\
J153231.80+420342.7 	&	300	&	0.21		&	0.878	&	1	\\
J082857.99+074255.7 	&	300	&	0.554	&	1.464	&	2	\\
J123605.45-014119.1 	&	330	&	0.211	&	1.012	&	2	\\
J124928.36+353926.8	&	380	&	0.527	&			&	2	\\
J133226.34+060627.3	&	420	&	0.207	&	2.664	&	2	\\
J134415.75+331719.1	&	730	&	0.686	&	1.041	&	2	\\
J144157.24+094859.1 	&	810	&	0.22		&			&	2	\\
J171544.02+600835.4	&	350	&	0.157	&	0.654	&	2	\\

\enddata

\tablecomments{Objects with equal-height components in the double-peaked \oiii\ line selected by eye (EHAGN), as described in Section \ref{sec:selection}. Note that the ratio of the fluxes in these objects is not necessarily near unity. Objects with equal-peaks selected by flux ratio are given in Table 2.}

\end{deluxetable}

%=============Table 2===============

\begin{deluxetable}{lcccc}
\label{tab:list}
\tablewidth{0pt}
\tablecaption{Equal Flux Double-Peaked AGN\label{t:tab2}}
\tablehead{
\colhead{SDSS Name} &
\colhead{Velocity Splitting} &
\colhead{$z_{\rm SDSS}$} &
\colhead{$F_r/F_b$} &
\colhead{Seyfert Type} \\
\colhead{ } &
\colhead{km/s} &
\colhead{ } &
\colhead{ } &
\colhead{ }}
\startdata
									
J081542.53+063522.9	&	280	&	0.244	&	0.8793	&	1	\\										
J090615.92+121845.6	&	330	&	0.644	&	1.2467	&	1	\\
J091649.41+000031.5 	&	310	&	0.222	&	0.8054	&	1	\\
J101241.20+215556.0	&	210	&	0.111	&	1.0270	&	1	\\
J120343.22+283557.8	&	850	&	0.374	&	0.8585	&	1	\\	
J121911.16+042905.9 	&	480	&	0.555	&	0.7735	&	1	\\	
J124813.82+362423.6	&	350	&	0.207	&	0.8273	&	1	\\	
J130724.08+460400.9	&	580	&	0.353	&	0.9400	&	1	\\	
J133455.24+612042.1	&	215	&	0.495	&	0.9733	&	1	\\
J145110.04+490813.5	&	220	&	0.156	&	0.8031	&	1	\\	
J145408.36+240521.3	&	300	&	0.535	&	0.7533	&	1	\\	
J151518.29+551535.3 	&	250	&	0.513	&	1.2140	&	1	\\	
J153231.80+420342.7 	&	300	&	0.210	&	0.8775	&	1	\\
J153423.19+540809.0	&	290	&	0.215	&	0.7596	&	1	\\	
J084049.46+272704.7 	&	580	&	0.136	&	0.8297	&	2	\\
J120526.04+321314.6 	&	670	&	0.485	&	0.7785	&	2	\\	
J123605.45-014119.1 	&	330	&	0.211	&	1.0123	&	2	\\
J134415.75+331719.1	&	730	&	0.686	&	1.0410	&	2	\\		
J140500.14+073014.1	&	340	&	0.135	&	0.8112	&	2	\\	
J140816.02+015528.3 	&	350	&	0.166	&	0.9778	&	2	\\
J151842.95+244026.0	&	280	&	0.561	&	1.0098	&	2	\\
J210449.13-000919.1 	&	380	&	0.135	&	0.9397	&	2	\\

\enddata

\tablecomments{Objects with equal-flux components in the double-peaked \oiii\ line (EFAGN), as described in Section \ref{sec:selection}.}

\end{deluxetable}

%===============Table 3=================

 \begin{deluxetable}{lccc}
 \label{tab:list}
 \tablewidth{0pt}
 \tablecaption{Line Ratios for Equal-Height DPAGN \label{t:tab3}}
 \tablehead{
 \colhead{SDSS Name} &
 \colhead{\oiii\ Flux} &
 \colhead{\nev\ Flux} &
 \colhead{\nev/\oiii}\\
 \colhead{ } &
 \colhead{ } &
 \colhead{ } &
 \colhead{ }}
 \startdata 
 
J113105.07+610405.1	 &	160		&	$<18$	&	$<0.110$\\	
J131018.47+250329.5      &	1419		&	80.0		&	0.056\\	
J151518.29+551535.3	 &	326		&	20.3		&	0.062\\
J153231.80+420342.7	&	1214		&	90.5		&	0.075\\

 \enddata
 
 \tablecomments{Line fluxes and ratios for eye-selected EHAGN.}
 
 \end{deluxetable}
\newpage

%=============Figure 1=========================

\newpage

%============= Figure 2 =============================

\end{document}